\documentclass{PoS}

\title{Present-Day Carbon Abundances\\ of Early-Type Stars}

\ShortTitle{Present-Day Carbon Abundances of Early-Type Stars}

\author{\speaker{M.~F.~Nieva}$^{ab}$ and N.~Przybilla$^a$\\
\llap{$^a$}Dr. Remeis-Sternwarte Bamberg, Sternwartstrasse 7, D-96049 Bamberg, Germany\\
\llap{$^b$}Observat\'orio Nacional, Rua G. Jos\'e Cristino 77, 20921-400, Rio de Janeiro, Brazil\\
E-mail: 
\email{nieva@sternwarte.uni-erlangen.de}}

\abstract{Carbon is one of the most abundant metals in the universe because of its synthesis in the fundamental 3$\alpha$ reaction. The knowledge of carbon abundances in different environments is one key ingredient to our understanding of stellar and galactochemical evolution. Studies of luminous OB-type stars allow us to address both topics even in galaxies beyond our own. Unfortunately the history of carbon abundance determinations from these objects in the last three decades is one of limited success. Analyses of the strong and weak line spectra of C\,{\sc ii} as well as C\,{\sc iii} tend to be largely discrepant.
We present results of quantitative spectral analyses based on a sophisticated model atom for non-LTE line-formation calculations of C\,{\sc ii-iv}. As a first application, carbon abundances in a sample of B-type dwarfs and giants in nearby associations and in the field are determined. Consistency is finally achieved for all measurable lines from the three ionization stages. This includes in particular the notorious C\,{\sc ii} $\lambda\lambda$4267 and 6578/6582~{\AA}~features which are highly important for abundance determinations of fast-rotating and extragalactic objects. The long-standing problem of carbon line formation can now be regarded as solved, with the previous difficulties related to the use of inaccurate atomic data and stellar parameters. A highly homogeneous and slightly sub-solar present-day carbon abundance from young stars in the solar vicinity of $\log$ C/H\,+\,12\,=\,8.32$\pm$0.04 is derived.
          }

\FullConference{International Symposium on Nuclear Astrophysics --- Nuclei in the Cosmos --- IX\\
		 June 25-30 2006\\
		 CERN, Geneva, Switzerland}

\begin{document}

\section {Introduction}

One of the most abundant metals in the universe is carbon, produced in the 3$\alpha$ reaction in massive stars and constituting the central building block of all organic chemistry.  
Abundances derived from luminous OB stars provide important
constrains on stellar and chemical evolution of our own and other galaxies. In
extragalactic applications (e.g. observations of main sequence stars in the Magellanic
Clouds, which have become feasible recently) astrophysicists desire to
analyse the {\em strongest} features in the metal line spectra, as these are the
only measurable at low signal-to-noise (S/N) and/or high projected rotational
velocities. In the case of ionized carbon these are the prominent lines of the two
multiplets C\,{\sc ii} $\lambda\lambda$4267.02/4267.27\,\AA~and
6578.03/6582.85\,{\AA}. These lines are unfortunately highly sensitive to
non-LTE effects, as well as to the choice of stellar atmospheric parameters.
So far, all studies from the literature failed to derive consistent abundances
from these lines and also to establish the C\,{\sc ii/iii} ionization equilibrium.

The present work aims to provide a solution to the classical non-LTE problem of
carbon abundance determinations in OB stars. A reliable C\,{\sc ii-iv} model atom
is developed and first applications on high-quality spectra are presented.
Besides great care in the selection of atomic data, special emphasis is also
given to an accurate atmospheric parameter determination, both in order to
minimise systematic uncertainties. The {\em entire} measurable C\,{\sc ii-iv} spectrum in the visual is investigated.

\section{Model Calculations}
A hybrid approach is used for the non-LTE line-formation computations. These are
based on line-blanketed plane-parallel, homogeneous and hydrostatic LTE model
atmospheres calculated with ATLAS9. Non-LTE synthetic
spectra are computed with recent versions of DETAIL and SURFACE. These codes solve the coupled radiative
transfer and statistical equilibrium equations and compute synthetic spectra
using refined line-broadening data, respectively.
Non-LTE line-formation computations for C\,{\sc ii/iii/iv}, hydrogen and He\,{\sc i/ii} are performed using state-of-the-art model atoms based on critically selected atomic data:~\cite{np06a},~\cite{np06c},~\cite{p05} and~\cite{pb04}.

\section{Observations and Analysis}
A first sample of six apparently slow-rotating Galactic B-type dwarfs and giants from OB associations and from the field in the solar vicinity is analysed. The observational database consists of very high S/N spectra obtained with FEROS on the 2.2m telescope at La Silla, Chile (ESO).

The stellar parameters are derived from application of an extensive iterative method
 resulting in simultaneous fits to all measurable H, He~\cite{np06b} and C\,{\sc ii-iv} lines~\cite{np06c}. 
 The iteration is performed on effective temperature $T_\mathrm{eff}$ and surface gravity $\log g$ (goal: to achieve ionization equilibrium) as well as the micro-, macroturbulent and projected rotational velocities (from carbon line profiles), He and C
abundances and different sets of atomic data. The final solution for each star is obtained when the statistical deviations of the averaged C abundance achieve their minimum. By application of the procedure to all programme stars it was possible to calibrate the C\,{\sc ii-iv} model atom for the entire parameter range (21\,500 $\le$ $T_\mathrm{eff}$ $\le$ 32\,000 K, 3.10 $\le$ $\log g$ $\le$ 4.30, dwarfs and giants) and simultaneously to obtain highly accurate C abundances for the sample.

\section{Sensitivity of Carbon Abundances to Parameter Variations}
\begin{figure}
\centering
\includegraphics[width=0.45\linewidth, angle=90]{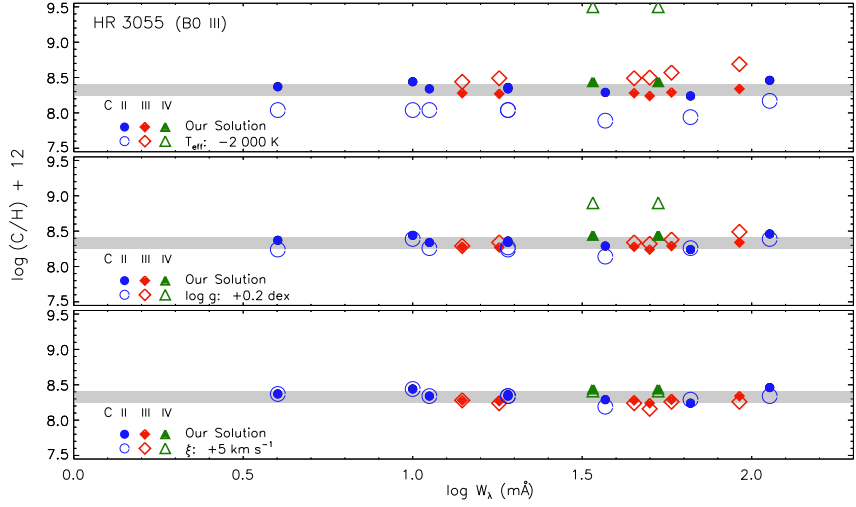}
\caption{Sensitivity of carbon abundances to typical parameter variations  found in the literature. Deviations from our solution for HR~3055 are shown for the 3 ionization stages.
The grey rectangles  correspond to 1$\sigma$-uncertainties of the stellar carbon abundance in our solution (upper left corner). The offsets of the parameters are also displayed.}
\label{parameters}
\end{figure}

Spectral lines of C, as well as many other metal lines, can be highly sensitive to atmospheric parameter variations. This behaviour is utilised to find the best solution for both stellar parameters and abundances.
In Fig.~\ref{parameters} non-LTE carbon abundances as a
function of the equivalent width $W_{\lambda}$ in HR~3055 are displayed for our solution (see final parameters in Fig.~\ref{results}) and variations in $T_\mathrm{eff}$, $\log g$ and microturbulence. The offsets of the parameters are typical for systematic discrepancies found in the literature.
 Note that the C\,{\sc iv} lines are extremely sensitive to changes in $T_\mathrm{eff}$ and $\log g$ at this temperatures (31\,200 K), with discrepancies up to $+$1.0\,dex in abundance. The C\,{\sc ii/iii} ionization equilibrium is also never established for a variation of these parameters (abundance changes up to $-$0.40\,dex for C\,{\sc ii} and up to $+$0.35\,dex for C\,{\sc iii} compared to our solution).
 An expected reduction of the abundances from strong lines is obtained for an increased microturbulence. Note that the scale in carbon abundance used is large (1.5 dex) and the variations of C abundances are significant relative to the high accuracy obtained in our solution.
The systematic variation of the total carbon abundance can reach 0.1\,dex and its statistical 1$\sigma$ value $\sim$ 0.4\,dex, when compared to our solution. These discrepancies increase when we use different sets of atomic input data. 
 
\section{Results for the Programme Stars}
\begin{figure}
\centering
\includegraphics[width=0.65\linewidth, angle=90]{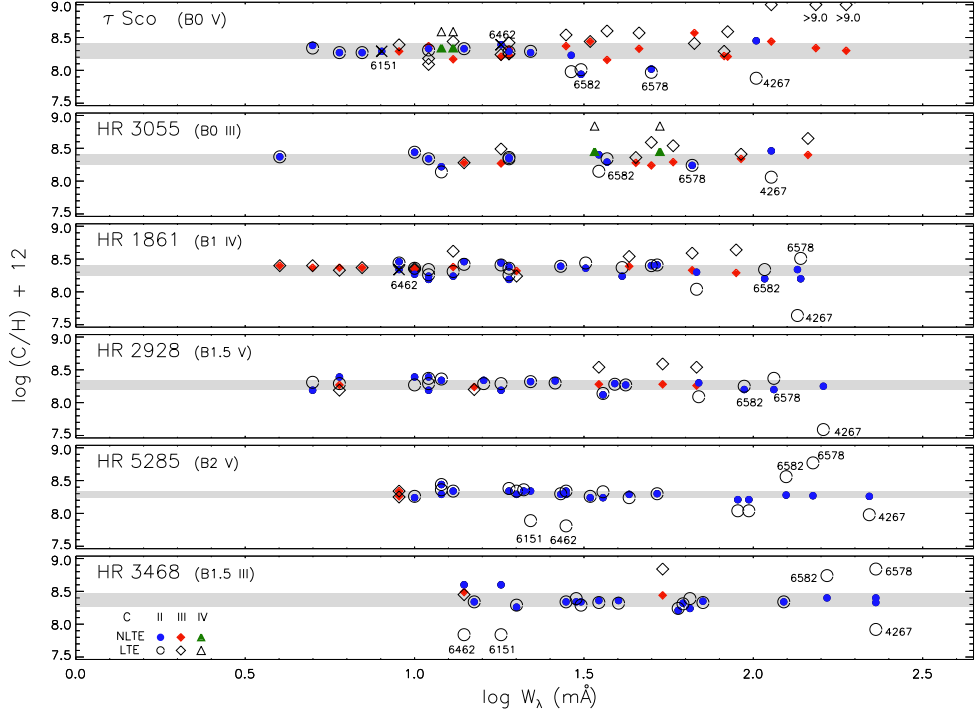}
\caption{
Non-LTE (filled symbols) and LTE (open symbols) abundances vs. equivalent width 
for C\,{\sc ii-iv} lines that could be measured in each spectrum. The ID of the stars and the derived mean carbon abundance is
 given in the upper left corner of each row, as well as derived basic stellar parameters ($T_\mathrm{eff}$, $\log g$, micro-, macroturbulent and projected rotational velocity). The grey rectangles
 correspond to 1$\sigma$-uncertainties of the stellar carbon abundance. Identification of lines with high sensitivity to non-LTE effects is displayed. Emission lines are marked by crosses (C\,{\sc ii} $\lambda\lambda$6151 and 6462\,\AA~in $\tau$\,Sco and $\lambda$6462\,\AA~in HR\,1861): LTE calculations are not able to reproduce them even qualitatively.}
\label{results}
\end{figure}

\begin{figure}
\centering
\includegraphics[width=0.7\linewidth]{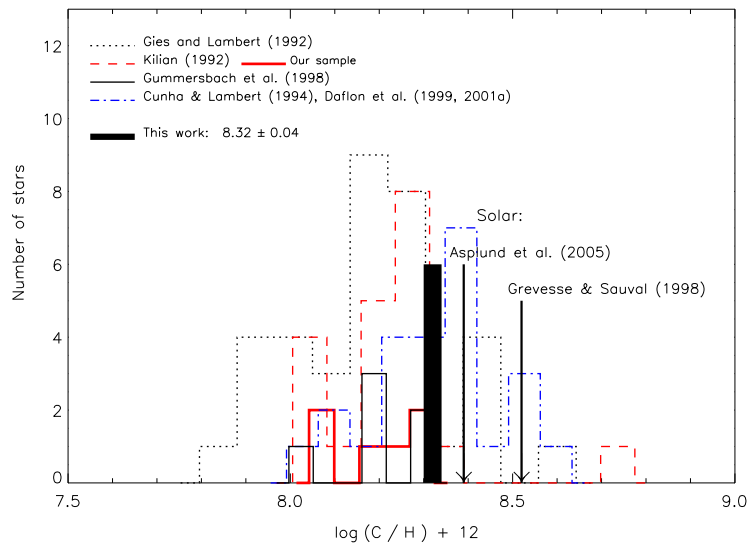}
\caption{Comparison of carbon abundances derived from our sample of early B-type stars in the solar vicinity with results from the literature and with recent solar values.}
\label{histogram}
\end{figure}

Non-LTE and LTE abundances vs. $W_{\lambda}$ for all individual lines are displayed in Fig.~\ref{results}, showing excellent consistency in non-LTE. 
 Identifications of the C\,{\sc ii} $\lambda\lambda$4267, 6578/83, 6151 and
6462\,\AA~lines are displayed. These lines are highly sensitive to
deviations from detailed equilibrium. The final parameters, as well as the carbon abundances, are also shown in Fig.~\ref{results}.
An immediate consequence of our careful analysis is a highly consistent value of carbon abundance free of systematic effects for all the stars of the sample.
In Fig.~\ref{histogram} we show a comparison of our first results for present-day carbon abundances of early B-type stars in the solar vicinity with results from the literature and with recent solar values (our stars coincide with 6 objects from~\cite{k92}). Despite our~--~so far~--~small sample, it is clear that the fact of avoiding systematic effects in the whole analysis leads to a highly consistent value of carbon abundance. An analysis of a larger sample has to be done in order to improve on the statistics. 

\section{Conclusions}
A highly consistent mean carbon abundance of
$\varepsilon$(C)\,$=$ 8.32$\pm$0.04 is derived from the six sample stars,
which provides a tight estimate to the present-day C abundance from young
stars in the solar vicinity. 
The atmospheric composition appears to be unaffected
by chemical mixing in the course of stellar evolution, i.e. we find no trend
of C abundances with evolutionary age. For comparison, adopting results from \cite{k92} one derives a mean $\varepsilon$(C)\,$=$\,8.19$\pm$0.12
from the same six stars, implying a considerable systematic shift and
a significantly increased statistical scatter. More objects need to be
analysed in order to confirm such homogeneous present-day (slightly) sub-solar carbon abundances -- considering as references $\varepsilon$(C)$_{\odot}$\,=\,8.39$\pm$0.05~\cite{ags05} or  $\varepsilon$(C)$_{\odot}$\,=\,8.52$\pm$0.06~\cite{gs98} -- in nearby associations (HR\,1861: Ori OB1; $\tau$\,Sco,
HR\,5285: Sco-Cen) and in the field (the other stars).
Despite the small sample size analysed so far, our highly accurate results indicate that the large scatter found for carbon abundances in early-type stars in previous work could be a consequence of systematic uncertainties introduced by the choice of inappropriate atomic data and stellar parameters.

\end{document}